\documentclass[twocolumn,showpacs,prb]{revtex4}
\usepackage{graphicx}
\usepackage{dcolumn}
\usepackage{amsmath}
\usepackage{latexsym}

\usepackage{hyperref}

\begin{document}
\title{\bf{Poincare gauge theory from higher derivative matter lagrangean}}
\author{ Pradip Mukherjee
$\,^{\rm a,b}$
}
\affiliation{$^{\rm a}$Department of Physics, Presidency College\\
86/1 College Street, Kolkata - 700 073, India}
\affiliation{$^{\rm b}$\tt mukhpradip@gmail.com}

\vspace{-0.5cm}

\date{}


\begin{abstract}Starting from matter lagrangean containing higher order derivative than the first, we construct the Poincare gauge theory by localising the Poincare symmetry of the matter theory. The construction is shown to follow the usual geometric procedure of gravitational coupling, thereby buttressing the geometric interpretation of the Poincare gauge theory.

\end{abstract}
\maketitle 
\section{Introduction}
Poincare Gauge Theory (PGT) is a celebrated framework where the gauge principle, so successful in the other branches of theoretical physics, is applied in the context of the theories of gravitation \cite{Utiyama:1956sy, Kibble:1961ba,Blagojevic:2002du}.
The edifice of PGT is constructed by localising the Poincare symmetry in Minkowski space. One starts with a matter theory invariant under global Poincare transformations. Naturally, this does not remain invariant when the parameters of the Poincare transformations are function of spacetime. The PGT emerges from attempts to modify the matter theory so that it becomes invariant under the local Poincare transformations. Compensating potentials are introduced in the process, the dynamics of which is provided by invariant densities constructed out of the field strengths obtained from the usual gauge theoretic procedure. The theory has been ubiquitous in classical gravity \cite{Blagojevic:2002du} as well as in its extension to noncommutative spacetime \cite{Chamseddine:2000si,Calmet:2005qm,Mukherjee:2006nd, Banerjee:2007th, rev}. 

  A very important aspect of the Poincare gauge theory is the correspondence of the gauge fields introduced here with the tetrad and spin - connection in Riemann -- Cartan spacetime.  The geometric interpretation of PGT, based on local Lorentz transformations (LLT) and general coordinate transformations or diffeomorphisms (diff), is a well -- known fact \cite{Blagojevic:2002du}. This has recently been
 explicitly demonstrated by compairing the transformations of the geometrical structures such as the metric and the  connection under LLT plus diff with those under the Poincare gauge transformations \cite{BGMR}. The usefulness of the PGT stems from this geometric interpretation i.e. the fact that theories invariant under local Poincare transformations can be viewed as invariant theories in curved spacetime.We can thus view the construction of PGT as a means of coupling the initial particle theory with gravity.

   The geometric interpretation of PGT is a cherished feature which is however more of an observed fact rather than a product of some underlying fundamental principle. Primarily, PGT originates from the localisation of the global Poincare symmetry of matter theories. It is a gauge theory in the Minkowski space and there is no {\it{a priori}} reason that the transformations of the gauge fields should mimick the transformations of the geometric structures in the curved space -- time. This correspondence is found {\it{en route}} \cite{BGMR} which enables one to identify the theory as carrying the symmetry of the curved space -- time. It is thus notable in this context that this procedure has been developed in the literature for lagrangeans containing first order derivatives only \cite{Blagojevic:2002du}.
The questions like whether the geometric interpretation will persist if we apply the gauge principle to a Poincare invariant theory containing higher derivative lagrangeans are nontrivial and demands thorough analysis.

 The construction of PGT starting from matter theory with higher derivative lagrangeans has not, so far, attracted much attention in the literature. On the other hand this problem has a direct bearing on the geometrical interpretation of PGT. If we look from the point of view of extending the gauge principle to theories with higher derivative lagrangeans we find that it is a complicated task which in general requires the introducion of new tensor gauge fields \cite{hama}. However, from the geometric point of view the coupling of lagrangean field theories with gravity is straight forward.Employing the principle of general covariance such a theory formulated in the Minkowski space is coupled to gravity by replacing the Minkowski metric by the metric of the curved space -- time and substituting ordinary derivatives by covariant derivatives \cite{Weinberg}. The localisation of Poincare innvariance of higher order matter theory i.e. the construction of PGT from such theories is thus worth investigating. The point is to verify whether the PGT construction corresponds to the geometric coupling when matter theories with higher derivative lagrangeans are taken as the starting point. This will indeed be an alternative independent check of the geometric interpretation of PGT. Moreover lagrangean theories with higher order derivatives are interesting in their own right. These have been discussed over a long period of time
\cite{lw,gitman1, gitman2, DN, HH, R, S, K, cl, AGMM}.
Such theories have appeared in different contexts: higher derivative terms naturally occur as quantum corrections to the lower order theories; various stringy models are shown to be equivalent to higher derivative theories;  the literature is rich in possible higher derivative theories of gravity which find application in quantum gravity \cite{BOS}.
It will thus be instructive  to pursue the PGT construction directly by localising the global Poincare symmetry of matter lagrangeans containing higher derivatives of the fields than the first. We propose to address this question in the present paper.

      The status of the PGT in the realm of gauge theory is indeed very special. The transformation of the potentials in the PGT obtained from canonical analysis is different from what is obtained from localisation of the Poincare symmetry. The two can be mapped only modulo the equations of motion \cite{BGMR, topo}. The construction of PGT from more general lagrangeans thus become more interesting in this perspective. Our results indeed vividly illustrates the unique role of PGT. At first terms upto the second derivatives are retained in the matter lagrangean. The question arises how to modify the second derivative terms so that the theory has local Poincare invariance. Taking the cue from the geometric interpretation of the usual theory, we substitute the second derivatives $\partial_k\partial_m$ by $\nabla_k\nabla_m$, where $\nabla$ is the covariant derivative. We find that the transformation of $\nabla_k\nabla_m\phi$ under local Poincare transformations {\it{is}} of the same form as that of $\partial_k\partial_m\phi$ under global Poincare transformations. 
No new gauge fields are required -- a remarkable fact from the point of view of extension of gauge principle to the higher derivative theories since it is known that in general such procedure involves the introduction of new (tensor) gauge fields \cite{hama}. The observation is in correspondence with the geometric interpretation of PGT \cite{Blagojevic:2002du, BGMR} and thus manifests the geometric significance of PGT in a more general context. 
 
  The above discussions apply to  lagrangean theories with second derivative terms. The purpose of the paper would remain unfullfilled if we do not demonstrate the validity of our construction for the  general case. We prove the validity of our construction in the general setting by induction. First we demonstrate that if our construction is valid for a theory containing $n$-th order derivatives then it is true for $n+1$ -th derivatives. But we have proved that the construction goes through for $n = 2$. By this inductive procedure the same geometric connection derived from a limited class of first order theories \cite{Utiyama:1956sy, Kibble:1961ba,Blagojevic:2002du}
is shown to be valid for the most general lagrangean theory.

    The organisation of the paper is as follows. In Section II the Poincare gauge theoretic construction from the usual lagrangean theories where the lagrangean contains only the first derivatives of the fields, is reviewed. While this material is standard it sets the stage for extension of the construction for more general theory. In the next section construction of Poincare gauge theory from higher derivative matter theory is presented. 
 The paper ends with a few concluding remarks in section IV. 

\section{Poincare gauge theory -- a review}
  We begin with the algorithm of constructing Poincare gauge theory (PGT) from a matter theory  when the lagrangean does not contain higher derivatives \cite{Blagojevic:2002du, Utiyama:1956sy, Kibble:1961ba}. The starting point of the PGT is the global Poincare transformations in the Minkowski space:
\begin{equation}
x^\mu \rightarrow x^\mu + \xi^\mu\label{GPT}
\end{equation}
where 
$$\xi^\mu = \theta^\mu_{\ \nu}x^\nu + \epsilon^\mu$$
 with both $\theta^{\mu\nu}$ and $\epsilon^\mu$ being infinitesmal constants and $\theta^{\mu\nu}$ antisymmetric.  We construct a local basis ${\bf{e}}_i$  at each spacetime point which are related to the coordinate basis ${\bf{e}}_\mu$ by 
$${\bf{e}}_i = \delta^\mu_i{\bf{e}}_\mu.$$
The fields will be assumed to refer to the local basis. The Latin indices refer to the local Lorentz frame and the Greek indices refer to the coordinate frame. The differentiation between the local and the coordinate basis appears to be a formal one here but it will be a necessity in curved space -- time.

   Now
consider a matter lagrangean 
 $${\cal{{L}}} = {\cal{{L}}}(\phi,\partial_{k}\phi)$$
 The corresponding action is 
\begin{equation}
I = \int d^4x {\cal{{L}}}(\phi,\partial_{k}\phi)\label{gaction}
\end{equation}
Under (\ref{GPT}) the action transforms as $$I \to I + \Delta I$$ where
\begin{equation}
\Delta I = \int d^4x \Delta {\cal{{L}}}(\phi,\partial_{k}\phi) \label{gactionvar}
\end{equation}
The variation
 $\Delta{\cal{L}}$ is given by
\begin{equation}
\Delta{\cal{L}} = \delta{\cal{L}} + \xi^\mu\partial_\mu{\cal{L}}+ \partial_\mu\xi^\mu{\cal{L}}
\label{oinvariance}
\end{equation}
where $\delta{\cal{L}}$
 is the form variation of the lagrangean
\begin{equation}
\delta{\cal{L}} = {\cal{L^\prime}}(x) - {\cal{L}}(x)
\end{equation}
${\cal{L^\prime}}$ being the transformed lagrangean {\footnote{The form variation of a quantity will always be denoted by the precedent $\delta$.}}. The condition for invariance of the theory is{\footnote{ In the most general case a total divergence may be added to the righthand side. However the condition (\ref{invariance}) is sufficiently general in (3+1) dimensions.}}
\begin{equation}
\Delta{\cal{L}} = 0
\label{invariance}
\end{equation}

  It is useful to scrutinise the above invariance condition with care. 
For global Poincare transformations 
\begin{equation}
\partial_\mu\xi^\mu=0\label{condition1}
\end{equation}
Also 
the field and its derivative transform as
\begin{eqnarray}
\delta\phi &=& \left(\frac{1}{2}\theta^{ij}\Sigma_{ij} - \xi^\mu\partial_\mu\right) \phi={\cal{P}}\phi\label{condition2a}\\
\delta\partial_k\phi &=& \left(\frac{1}{2}\theta^{ij}\Sigma_{ij} - \xi^\mu\partial_\mu\right)\partial_k\phi + \theta_k^{\ j}\partial_j\phi \nonumber\\
&=& {\cal{P}}\partial_k\phi + \theta_k^{\ j}\partial_j\phi\label{condition2b}
\end{eqnarray}
$\Sigma_{ij}$ are the Lorentz spin matrices, the form of which depend on the particular representation to which $\phi$ belong. These are matrices with constant elements that satisfy the Lorentz algebra
\begin{equation}
\left[\Sigma_{ij},\Sigma_{kl}\right] = \eta_{il}\Sigma_{jk} -
      \eta_{ik}\Sigma_{jl} + \eta_{jk} \Sigma_{il} - \eta_{jl}\Sigma_{ik}\label{algebra}  
\end{equation}
The equations (\ref{condition1}), (\ref{condition2a})and (\ref{condition2b}) are instrumental for the invariance condition (\ref{invariance}) to be satisfied.  The form variation $\delta{\cal{L}}$ is explicitly given by
\begin{equation}
\label{lvariation}
\delta{\cal{L}}= \frac{\delta {\cal{L}}}{\delta\phi}\delta\phi + \frac{\delta {\cal{L}}}{\delta\partial_k\phi}\delta\partial_k\phi
\end{equation}
and it is the form of variations given by (\ref{condition2a}), (\ref{condition2b})and the condition (\ref{condition1}) which lead to the invariance of the theory. 

  When the Poincare symmetry (\ref{GPT}) is assumed to be a local symmetry the parameters $\theta$ and $\epsilon$ are no longer constants but becomes function of space -- time coordinates.  
 Here, it is advantageous to take the  parameters  $\xi^\mu = \theta^\mu_{\ \,\nu} x^\nu +  \epsilon^\mu$ and $\theta^{ij}$ as the independent parameters. It is natural that the action (\ref{gaction}) which was invariant under (\ref{GPT}) with constant parameters ( global Poincare transformations ) will cease to remain invariant when the parameters become function of spacetime ( local Poincare transformations ).
The reasons are:

(i) equation (\ref{condition1}) is no longer true and

(ii) though the fields transform as equation (\ref{condition2a}) :

\begin{eqnarray}
\delta\phi &=& \left(\frac{1}{2}\theta^{ij}\Sigma_{ij} - \xi^\mu\partial_\mu\right) \phi\nonumber
           ={\cal{P}}\phi
\end{eqnarray}
their derivatives $\partial_k\phi$ transform as
\begin{eqnarray}
\delta\partial_k\phi &=& \left(\frac{1}{2}\theta^{ij}\Sigma_{ij} - \xi^\mu\partial_\mu\right)\partial_k\phi - \partial_k\xi^\nu\partial_\nu\phi + \frac{1}{2}\partial_k\theta^{ij}\Sigma_{ij}\phi\nonumber\\
           &=& {\cal{P}}\partial_k\phi - \partial_k\xi^\nu\partial_\nu\phi + \frac{1}{2}\partial_k\theta^{ij}\Sigma_{ij}\phi\label{condnew} 
\end{eqnarray}
which is different from their counterpart in equation (\ref{condition2b}).

 To modify the matter action so as to be invariant under the {\it{local}} Poincare transformations, one has to remeady the above departures. The first thing is to replace the ordinary derivative $\partial_k\phi$ by some covariant derivative $\nabla_k\phi$ which will transform  as in (\ref{condition2b}).
 This is done in two steps:

(i) In the first step the $\theta$ - covariant derivative $\nabla_\mu$ is introduced which eliminates the $\partial_\mu\theta^{ij}$ term from (\ref{condnew}). We define  $\nabla_\mu$ as
\begin{equation}
\nabla_\mu = \partial_{\mu} + \frac{1}{2}\omega^{ij}_{\ \ \mu}\Sigma_{ij}\label{nablamu}
\end{equation}
where $\omega^{ij}_{\ \ \mu}$ are the `gauge potentials'.The required transformation of $\nabla_\mu\phi$ is
\begin{equation}
\delta\nabla_\mu\phi = {\cal{P}}\nabla_{\mu}\phi -\partial_\mu \xi^{\nu}\nabla_\nu\phi
\end{equation}
The transformation of the 'gauge Field' 
$\omega^{ij}_{\ \ \mu}$ is determined from this requirement.

(ii) In the next step the covariant derivative in the local frame is constructed as 
\begin{equation}
\nabla_k = b_k^{\ \mu}\nabla_\mu\label{covder}
\end{equation}
where $b_k^{\ \mu}$ is another compensating field. For later convenience we define
$b^k_{\ \mu}$ as the inverse to $b_k^{\ \mu}$. 
If the fields $b^i_{\ \mu}$ and $\omega^{ij}_{\ \ \mu}$ transform as 
\begin{eqnarray}
\delta b^i_{\ \mu} &=& \theta^i_{\ k} b^k_{\ \mu} - \partial_\mu\xi^\rho b^i_{\ \rho} - \xi^{\rho}\partial_{\rho} b^i_{\ \mu}\nonumber\\
\delta \omega^{ij}_{\ \ \mu} &=& \theta^i_{\ k} \omega^{kj}_{\ \ \mu} + \theta^j_{\ k} \omega^{ik}_{\ \ \mu} - \partial_\mu\theta^{ij} - \partial_\mu\xi^\rho \omega^{ij}_{\ \ \rho}\nonumber
\\
 &-& \xi^{\rho}\partial_{\rho}\omega^{ij}_{\ \ \mu}\label{fieldtrans}
\end{eqnarray} 
then  $\nabla_k\phi$
 transforms as 
\begin{equation}
\delta\nabla_k\phi = {\cal{P}}\nabla_k\phi + \theta_k^{\ j}\nabla_j\phi\label{C}
\end{equation}
This transformation rule is formally identical with
(\ref{condition2b}). So we get the required transformation.
  The matter lagrangian density ${\cal{L}} = {\cal{L}}(\phi,\partial_{k}\phi)$ which was invariant under global Poincare transformations is converted to an invariant density ${\cal{\tilde{L}}}$ under local Poincare transformations by replacing the ordinary derivative $\partial_{k}$ by the covariant derivative $\nabla_k $ i.e. 
$${\cal{\tilde{L}}} = {\cal{\tilde{L}}}(\phi,\nabla_{k}\phi).$$ 
The departure from equation (\ref{condition1}) can be accounted for by altering the measure of spacetime integration suitably. An invariant action is now constructed as 
$$I = \int d^4x b{\cal{\tilde{L}}}(\phi,\nabla_{k}\phi)$$
 where $b = {\rm det}\,b^i_{\ \mu}.$ The invariance of this action is ensured by the transformations (\ref{fieldtrans}) of the 'potentials'.

      At this point a very interesting property of the above construction should be noted. The transformations (\ref{fieldtrans}) comprise the Poincare gauge transformations. Their structure suggests a geometric interpretation. The basic fields $b^i_{\ \mu}$ and $\omega^{ij}_{\ \ \mu}$ mimic the tetrad and the spin connection in curved spacetime. The most general invariance group in curved spacetime consists of the LLT plus diff. Observe in (\ref{fieldtrans}), that the Latin indices transform as under LLT with parameters $\theta^{ij}$, and the Greek indices transform as under diff with parameters $\xi^\mu$. This suggests a correspondence between the Poincare gauge transformations with the geometric transformations of the curved spacetime. That this correspondence is an equivalence, has recently been demonstrated explicitly \cite{BGMR}. The geometric interpretation of PGT is a crucial step. It emerges from the application of the gauge principle. At this point it is noticable that the construction is limited to lagrangeans which contain only the first derivative of the fields. The question what happens for higher derivative theories begs to be addressed. One wonders whether the geometric interpretation will still hold? Otherwise the geometric connection will be episodic. The question of higher derivative theories will thus provide an independent check of the geometric connection.

 With the question of higher derivative theories pending for the time being, we come back to the construction of PGT in the usual context. Corresponding to the basic fields $b^i_{\ \mu}$ and $\omega^{ij}_{\ \ \mu}$ the Lorentz field strength $R^{ij}_{\ \ \mu\nu}$ and the translation field strength $T_{i\mu\nu}$ are obtained following the usual procedure in gauge theory. The commutator of two $\theta$-covariant derivatives gives $R^{ij}_{\ \ \mu\nu}$
$$ \left[\nabla_\mu,\nabla_\nu\right]\phi = \frac{1}{2}R^{ij}_{\ \ \mu\nu}\Sigma_{ij}\phi$$
whereas the commutator of two $\nabla_k$ derivatives furnish the additional fields $T_{i\mu\nu}$ as
\begin{equation}
 \left[\nabla_k,\nabla_l\right]\phi = \frac{1}{2} ~b_k^{\ \mu} b_l^{\ \nu} ~R^{ij}_{\ \ \mu\nu}\Sigma_{ij}\phi - b_k^{\ \mu}b_l^{\ \nu}~T^i_{\ \mu\nu}\nabla_i\phi\label{commutator}
\end{equation}
These defining equations give the following expressions for the field-strengths
\begin{eqnarray}
\label{pgt}
T^i_{\ \mu\nu} &=& \partial_\mu b^i_{\ \nu} + \omega^{i}_{\ \, k\mu} b^k_{\ \nu} - \partial_\nu b^i_{\ \mu} - \omega^{i}_{\ \,k\nu} b^k_{\ \mu}\nonumber\\
R^{ij}_{\ \ \mu\nu} &=& \partial_\mu \omega^{ij}_{\ \ \nu} - \partial_\nu \omega^{ij}_{\ \ \mu} + \omega^i_{\ k\mu}\omega^{kj}_{\ \ \nu} - \omega^i_{\ k\nu}\omega^{kj}_{\ \ \mu}.
\end{eqnarray}

  We have already observed the connection of Poincare gauge symmetry with LLT plus diff invariances in curved spacetime. This connection may further be pursued at the level of the field strengths. From the point of view of PGT the transformations of the Lorentz field strength $R^{ij}_{\ \ \mu\nu}$ and the translation field strength $T^i_{\ \mu\nu}$ can easily be obtained from (\ref{fieldtrans}) and (\ref{pgt}),
\begin{eqnarray}
\label{lorentztrans}
\delta R^{ij}_{\ \ \mu\nu} = \theta^i_{\ k}\,R^{kj}_{\ \ \mu\nu} &+&\theta ^j_{\ k}\,R^{ik}_{\ \ \mu\nu} -\partial_{\mu}\xi^{\rho}\,R^{ij}_{\ \ \rho\nu}\nonumber\\&-& \partial_{\nu}\xi^{\rho}\,R^{ij}_{\ \ \mu\rho} -\xi^{\rho}\,\partial_{\rho}R^{ij}_{\ \ \mu\nu}
\end{eqnarray}
and
\begin{eqnarray}
\label{transtrans}
\delta T^{i}_{\ \mu\nu} &=& \theta ^i_{\ k}\,T^{k}_{\ \ \mu\nu} -\partial_{\mu}\xi^{\rho}\,T^{i}_{\ \rho\nu} - \partial_{\nu}\xi^{\rho}\,T^{i}_{\ \mu\rho}\nonumber\\&-& \xi^{\rho}\,\partial_{\rho}T^{i}_{\ \mu\nu}
\end{eqnarray}
As noted earlier in the context of the transformations (\ref{fieldtrans}), the Latin indices transform as under LLT with parameters $\theta^{ij}$ and the Greek indices transform as under diff with parameters $\xi^\mu$. We get the expected transformations under LLT and diff.
 
  The locally Poincare invariant theory is constructed in the Minkowski space and has been developed as a gauge theory. Using the geometric interpretation, the Lorentz field strength $R^{ij}_{\ \ \mu\nu}$ and the translation field strength $T_{i\mu\nu}$, may be identified with the Riemann tensor and the torsion respectively. Using these basic structures, gravity can be formulated in the framework of PGT. The gravitational dynamics follows in general from combination of the Riemann tensor and the torsion
which are invariant under (\ref{fieldtrans}). 
 The procedure of obtaining the most general gravitational dynamics have been elaborately studied in the literature \cite{hayashi} but details of these are not required in the present discussion. We only note the crucial role of the geometric correspondence in casting PGT as theory of gravity.

\section{Localisation of Poincare symmetry of theories with higher order derivative}
 So far we were reviewing the methodology of coupling lagrangean field theories in the framework of PGT when the lagrangean of the theory contains only first order derivatives. While this material is standard, it paves the way to further generalisation. We will attempt to extend the same basic mechanism for theories with higher order derivative. For simplicity we will begin with theories inclusive of second order derivative . Note that the extension of gauge principle to higher derivative theories is known to be associated with the introduction of new gauge fields \cite{hama}.It is thus remarkable that we do not introduce any new gauge potential. 

        We consider theories whose action is given by 
\begin{equation}
I = \int d^4x {\cal{{L}}}(\phi,\partial_{k}\phi,\partial_k\partial_l\phi)\label{ghaction}
\end{equation}
It is assumed that the theory is invariant under global Poincare transformations. This means that under the transformation (\ref{GPT})
\begin{equation}
\delta I = \int d^4x \Delta{\cal{L}} = 0
\label{hinvariance}
\end{equation} where
\begin{equation}
\Delta{\cal{L}} = \delta{\cal{L}} + \xi^\mu\partial_\mu{\cal{L}}+ \partial_\mu\xi^\mu{\cal{L}}
\label{dinvariance}
\end{equation}
with
\begin{equation}
\label{hlvariation}
\delta{\cal{L}}= \frac{\delta {\cal{L}}}{\delta\phi}\delta\phi + \frac{\delta {\cal{L}}}{\delta\partial_k\phi}\delta\partial_k\phi + \frac{\delta {\cal{L}}}{\delta\partial_k\partial_l\phi}\delta\partial_k\partial_l\phi 
\end{equation} 

  Our concern is to modify the action (\ref{ghaction}) in such a way so that it is invariant under local Poincare transformation. The catch is in (\ref{hlvariation}).  Note the change in (\ref{hlvariation}) compared with (\ref{lvariation}).
Due to the presence of the second derivative in the lagrangean an extra term has appeared in (\ref{hlvariation}) as compared with (\ref{lvariation}). Let us calculate the form variation of $\partial_k\partial_l\phi$ under global Poincare transformation (\ref{GPT}). This is
\begin{equation}
\delta\partial_k\partial_l\phi = {\cal{P}}\partial_k\partial_l\phi + \theta_k^{\ m}\partial_m\partial_l\phi +\theta_l^{\ m}\partial_k\partial_m\phi \label{newvariation}
\end{equation}
To get a theory of the form (\ref{ghaction}) invariant under local Poincare gauge transformation one has to look for pieces that transform as (\ref{newvariation}). A possible candidate is 
$\nabla_k\nabla_l\phi$ but one has to be sure about its transformation properties.

           We will now obtain the transformation of
 $\nabla_k\nabla_l\phi$ under local Poincare transformation. Substituting the expression of $\nabla_k$ from (\ref{covder}) we can write
\begin{equation}
\nabla_k\nabla_l\phi = b_k^{\ \mu}\nabla_\mu\left(b_l^{\ \nu}\nabla_\nu\phi\right)\label{dpartial} 
\end{equation} 
The covariant derivative operator $\nabla_\mu$ operates everything on the right of it. The action on $b_l^{\ \nu}$ is computed from the same general expression (\ref{nablamu}) where $\Sigma_{ij}$ is now given by the appropriate vector representation
\begin{equation}
\left[\Sigma_{ij}\right]_{\ m }^n = \left(\eta_{jm}\delta^n_i -
\eta_{im}\delta^n_j\right)
\end{equation}
Using this we get from (\ref{dpartial})
\begin{equation}
\nabla_k\nabla_l\phi = b_k^{\ \mu}\left(\partial_\mu b_l^{\ \nu} - \omega^s_{\ l\mu}b_s^{\ \nu}\right )\nabla_\nu\phi + b_k^{\ \mu}b_l^{\ \nu}\nabla_\mu \nabla_\nu\phi \label{dnabla} 
\end{equation}
The form variation of $\nabla_k\nabla_l\phi$ is to be worked out from (\ref{dnabla}). 
A straightforward way is to expand (\ref{dnabla}) using (\ref{nablamu}) and then take the variation. After a long calculation we get
the variation of $\nabla_k\nabla_l\phi$ as
\begin{eqnarray}
\delta\nabla_k\nabla_l\phi = {\cal{P}}\nabla_k\nabla_l\phi &+&\theta_k^m\nabla_m\nabla_l\phi +\theta_k^m\nabla_k\nabla_m\phi \label{newvariation2}
\end{eqnarray}
  The transformation (\ref{newvariation2}) is quite of the required form (\ref{newvariation}).

   The transformation relations obtained in (\ref{newvariation2}) allows us to check the internal consistency of the Poincare gauge theory in the following way. Using (\ref{newvariation2}) we can calculate
$\delta\left[\nabla_k,\nabla_l\phi\right]$ as
\begin{eqnarray}
\label{check1}
\delta\left[\nabla_k,\nabla_l\right]\phi &=&{\cal{P}}\left[\nabla_k,\nabla_l\right]\phi +
           \theta_k^{\ m}\left[\nabla_m,\nabla_l\right]\phi \nonumber\\&+&\theta_k^{\ m}
\left[\nabla_k,\nabla_m\right]\phi 
\end{eqnarray}

 We have calculated the variation of $\left[\nabla_k,\nabla_l\right]\phi$ using our expression (\ref{newvariation2}) for $\delta\nabla_k\nabla_l\phi$.
Now this  can also be computed in an alternative way. $\left[\nabla_k,\nabla_l\right]\phi$ is given by (\ref{commutator}) where it is expressed in terms of the field strengths which have known transformation properties ( equations (\ref{lorentztrans}) and (\ref{transtrans})). Using (\ref{commutator}) we can write
\begin{equation}
\left[\nabla_k,\nabla_l\right]\phi = \frac{1}{2} R^{ij}_{\ \ kl}\Sigma_{ij}\phi  -  T^s_{\ kl}\nabla_s\phi \label{comnew}
\end{equation}
where we have used the definitions
\begin{eqnarray}
R^{ij}_{\ \ kl} &=& b_k^{\ \mu}b_l^{\ \nu}R^{ij}_{\ \ \mu\nu}\nonumber\\
T^i_{\ kl}  &=& b_k^{\ \mu}b_l^{\ \nu}T^i_{\ \mu\nu}\label{A}
\end{eqnarray}
The form variation of these fields may be worked out using equations (\ref{fieldtrans}), (\ref{lorentztrans})and (\ref{transtrans}). First we express
\begin{eqnarray}
\delta\left[\nabla_k,\nabla_l\right]\phi &=& \frac{1}{2}\delta R^{ij}_{\ \ kl}\Sigma_{ij}\phi + \frac{1}{2}R^{ij}_{\ \ kl}\Sigma_{ij}\delta\phi\nonumber\\ &-&\delta T^s_{\ kl}\nabla_s\phi
- T^s_{\ kl}\delta\nabla_s\phi\label{B} 
\end{eqnarray}
Substituting $\delta R^{ij}_{kl}$ and $\delta T^s_{kl}$ working from (\ref{A}) in(\ref{B}) we get
\begin{eqnarray}
\label{check2}
\delta\left[\nabla_k,\nabla_l\right]\phi = {\cal{P}}\left[\nabla_k,\nabla_l\right]\phi &+& 
           \theta_k^{\ m}\left[\nabla_m,\nabla_l\right]\phi \nonumber\\&+&\theta_k^{\ m}
\left[\nabla_k,\nabla_m\right]\phi 
\end{eqnarray}
The variation (\ref{newvariation2}) is thus found to fit in the general scheme. From the geometric point of view the variation (\ref{newvariation2}) is what it should be under diff and LLT. We thus continue to note the correspondence of the geometric and the gauge structures.




Coming back to the variation of $\nabla_k\nabla_l\phi$ obtained in (\ref{newvariation2}) we
note that it is formally identical with the transformation of $\delta\partial_k\partial_l\phi$
given by equation (\ref{newvariation}). The situation is then similar to what we observed for the usual theory. We have a theory (\ref{ghaction}) which is invariant under global Poincare transformation. The invariance is ensured by the specific forms of the variations of $\phi$, $\partial_k\phi$ and $\partial_k\partial_l\phi$ given by equations (\ref{condition2a}),(\ref{condition2b}) and (\ref{newvariation}). When the global Poincare transformations are made local, all these pieces naturally do not satisfy the same forms of variations but we can construct a covariant derivative $\nabla_k\phi$ so that $\nabla_k\phi$ and $\nabla_k\nabla_l\phi$ transform respectively as (\ref{condition2b}) and (\ref{newvariation}). 

           We can now form a prescription for the localisation of the Poincare invariance of higher order theories. It is simple: replace $\partial_k\phi$ by $\nabla_k\phi$ and $\partial_k\partial_l\phi$ by $\nabla_k\nabla_l\phi$ in a theory which was invariant under global Poincare transformations and a theory invariant under local Poincare transformations will be obtained. No new gauge field is required. This is again equivalent to the formulation of the theory in curved space -- time.This illustrates the correspondence of PGT with theories in curved space -- time from a more general perspective than was considered earlier \cite{Blagojevic:2002du,BGMR}. Giving independent dynamics for the gravitational fields complete the formulation of the theory under gravity. 

          Before passing to the next part of the analysis a particular feature is worth mentioning. Since $\nabla_k$ does not commute with $\nabla_l$ there is an ordering ambiguity. Thus , for instance, we could replace $\partial_k\partial_l\phi$ by
${\cal{D}}_{kl}\phi$ given by
\begin{equation}
{\cal{D}}_{kl}\phi = \frac{1}{2}\left(\nabla_k\nabla_l\phi + \nabla_l\nabla_k\phi\right)
\end{equation}
Indeed,because of (\ref{newvariation2})
\begin{eqnarray}
\delta {\cal{D}}_{kl}\phi &=& \frac{1}{2}[{\cal{P}}\nabla_k\nabla_l\phi + \theta_k^{\ m}\nabla_m\nabla_l\phi +\theta_l^{\ m}\nabla_k\nabla_m\phi\nonumber\\ &+& {\cal{P}}\nabla_l\nabla_k\phi + \theta_l^{\ m}\nabla_m\nabla_k\phi +\theta_k^{\ m}\nabla_l\nabla_m\phi ]
\end{eqnarray}
Simplifying, we get
\begin{equation}
\delta {\cal{D}}_{kl}\phi = {\cal{P}}{\cal{D}}_{kl}\phi + \theta_k^{\ m}{\cal{D}}_{ml}\phi +\theta_l^{\ m}{\cal{D}}_{km}\phi
\end{equation}
We find that the form variation of ${\cal{D}}_{kl}\phi$ is again the same as (\ref{newvariation}). This shows that the prescription of  replacement of $\partial_k\partial_l\phi$ is not unique.
This nonuniqueness is indeed due to noncommutativity of the covariant derivative $\nabla_k$.

         We have so far considered upto second derivatives of the fields in the lagrangean. The generalisation to arbitrary order may  be obtained by induction. To build the induction process it will be advantageous to redo the calculation of $\delta\nabla_k\nabla_l\phi$ in another way where the variation of $\nabla_k\phi$, given by (\ref{C}) will be directly used. Using (\ref{dnabla}) we can write
\begin{equation}
\nabla_k\nabla_l\phi = b_k^{\ \mu}\left(\partial_\mu + \frac{1}{2}\omega^{ij}_{\ \ \mu} \Sigma_{ij}\right)\nabla_l\phi - b_k^{\ \mu}\omega^{s}_{\ l\mu}\nabla_s\phi\label{in}
\end{equation}
In the above expression the matrix $\Sigma_{ij}$ is in the representation of $\phi$ and acts on it. Taking form variation
we get
\begin{equation}
\delta\left(\nabla_k\nabla_l\phi\right) = \delta\left(b_k^{\ \mu}\left(\partial_\mu + \frac{1}{2}\omega^{ij}_{\ \ \mu} \Sigma_{ij}\right)\nabla_l\phi\right) - \delta\left(b_k^{\ \mu}\omega^{s}_{\ l\mu}\nabla_s\phi\right)\label{nnabla}
\end{equation}
Now
\begin{eqnarray}
&&\delta\left(b_k^{\ \mu}\left(\partial_\mu + \frac{1}{2}\omega^{ij}_{\ \ \mu} \Sigma_{ij}\nabla_l\phi\right)\right)\nonumber\\ &=& \left(\delta b_k^{\ \mu}\right)\partial_\mu\nabla_l\phi + b_k^{\ \mu}\delta\left(\partial_\mu\nabla_l\phi\right)\nonumber\\&+&\left(\delta b_k^{\ \mu}\right)\frac{1}{2}\omega^{ij}_{\ \ \mu}\Sigma_{ij}\nabla_l\phi \nonumber\\&+& b_k^{\ \mu}\delta\left(\frac{1}{2}\omega^{ij}_{\ \ \mu}\Sigma_{ij}\nabla_l\phi\right)\label{dpartialvar} 
\end{eqnarray}
To compute 
r.h.s of (\ref{dpartialvar}) we substitute $\delta{b_k^\mu}$
and $\delta\omega^{ij}_{\ \ \mu}$ from (\ref{fieldtrans}). Also note that
\begin{equation}
\delta\left(\partial_\mu\nabla_l\phi\right) = \partial_\mu\left({\cal{P}}\nabla_l\phi + \theta_l^{\ m}\nabla_m\phi\right)
\end{equation}
which is obtained from (\ref{C}).
Substituting all these in (\ref{dpartialvar}) we get
\begin{eqnarray}
\delta\left(b_k^{\ \mu}\left(\partial_\mu \right.\right.&+&\left.\left. \frac{1}{2}\omega^{ij}_{\ \ \mu} \Sigma_{ij}\right)\nabla_l\phi\right)\nonumber\\  &=&{\cal{P}}\left(b_k^{\ \mu}\left(\partial_\mu + \frac{1}{2}\omega^{ij}_{\ \ \mu} \Sigma_{ij}\nabla_l\right)\phi\right)\nonumber\\  &+&\theta_k^m \left(b_m^{\ \mu}\left(\partial_\mu + \frac{1}{2}\omega^{ij}_{\ \ \mu} \Sigma_{ij}\right)\nabla_l\phi\right)\nonumber\\ &+& \theta_l^m \left(b_k^{\ \mu}\left(\partial_\mu + \frac{1}{2}\omega^{ij}_{\ \ \mu} \Sigma_{ij}\right)\nabla_m\phi\right)\nonumber\\&+& b_k^{\mu}\partial_\mu\theta_l^{\ m}\nabla_m\phi\label{nnabla1}
\end{eqnarray}
Similarly we find 
\begin{eqnarray}
\delta\left(b_k^{\ \mu}\omega^{s}_{\ l\mu}\nabla_s\phi\right)
&=&{\cal{P}}\left(b_k^{\ \mu}\omega^{s}_{\ l\mu}\nabla_s\phi\right)+ \theta_k^{\ m}\left(b_m^{\ \mu}\omega^{s}_{\ l\mu}\nabla_s\phi\right)\nonumber\\ &+& \theta_l^{\ m}\left(b_k^{\ \mu}\omega^{s}_{\ m\mu}\nabla_s\phi\right) \nonumber\\&+& b_k^{\mu}\partial_\mu\theta_l^{\ m}\nabla_m\phi\label{nnabla2}
\end{eqnarray}
Substituting (\ref{nnabla1}) and (\ref{nnabla2}) in (\ref{nnabla}) we get the form of $\delta\left(\nabla_k\nabla_l\phi\right)$ which of course agrees with (\ref{newvariation2}). The merit of the expansion (\ref{in}) is that it allows us to find $\delta\left(\nabla_k\nabla_l\phi\right)$ from $\delta\nabla_k\phi$ using the transformations of the basic fields.
This expansion can be generalised for n-th order covariant derivative as
\begin{eqnarray}
\nabla_{k_1}\nabla_{k_2} ....\nabla_{k_n}\phi &=& b_{k_1}^{\ \mu}\left(\partial_\mu + \frac{1}{2}\omega^{ij}_{\ \ \mu} \Sigma_{ij}\right)\nabla_{k_2}....\nabla_{k_n}\phi \nonumber\\&-& b_{k_1}^{\ \mu}\omega^{s}_{\ k_2\mu}\nabla_s\nabla_{k_3}...\nabla_{k_n}\phi\nonumber\\&-&  b_{k_1}^{\ \mu}\omega^{s}_{\ k_3\mu}\nabla_{k_2}\nabla_s...\nabla_{k_n}\phi....\nonumber\\&-& b_{k_1}^{\ \mu}\omega^{s}_{\ k_n\mu}\nabla_{k_2}\nabla_{k_3}...\nabla_{s}\phi\label{in1}
\end{eqnarray}
We will use (\ref{in1}) in the following for demonstration of our construction for Lagrangeans containing derivatives of arbitrary order.

 Assume that the construction is valid for a theory which contains upto $n$-th order derivative.This means that we replaced 
$\partial_{k_1}..... \partial_{k_n}\phi$ by $\nabla_{k_1}..... \nabla_{k_n}\phi$ such that

\begin{eqnarray}
\delta\nabla_{k_1}..\nabla_{k_n}\phi &=& {\cal{P}}\nabla_{k_1}..\nabla_{k_n}\phi \nonumber\\ &+&\theta_{k_1}^{\ m}\nabla_{m}..\nabla_{k_n}\phi +..\theta_{k_n}^{\ m}\nabla_{k_1}..\nabla_{m}\phi 
\label{newvariation32}
\end{eqnarray}
which ensures local Poincare invariance. We assume that equation (\ref{newvariation32}) holds for some n. Assume now that we have a theory containing the $n+1$-th derivative. We will show that 
 $\delta\nabla_{k_1}\nabla_{k_2}....\nabla_{k_{n+1}}\phi$ also satisfies equation (\ref{newvariation32}) for $n = n+1$. 
Indeed, using (\ref{in1}) we get
\begin{eqnarray}
&&\delta\left(\nabla_{k_1}\nabla_{k_2} ....\nabla_{k_{n+1}}\phi\right)\nonumber\\ &=& \delta\left(b_{k_1}^{\ \mu}\left(\partial_\mu + \frac{1}{2}\omega^{ij}_{\ \ \mu} \Sigma_{ij}\right)\nabla_{k_2}....\nabla_{k_{n+1}}\phi\right) \nonumber\\&-& \delta\left(b_{k_1}^{\ \mu}\omega^{s}_{\ k_2\mu}\nabla_s\nabla_{k_3}...\nabla_{k_{n+1}}\phi\right)\nonumber\\&-&  \delta\left(b_{k_1}^{\ \mu}\omega^{s}_{\ k_3\mu}\nabla_{k_2}\nabla_s...\nabla_{k_{n+1}}\phi\right)....\nonumber\\&-& \delta\left(b_{k_1}^{\ \mu}\omega^{s}_{\ k_{n+1}\mu}\nabla_{k_2}\nabla_{k_3}...\nabla_{s}\phi\right)\label{in11}
\end{eqnarray}
Compare this expression with that given by (\ref{nnabla}). There we found the transformations of the first term generates such pieces which are either cancelled by the appropriate terms coming from the second or couple with them so that the left hand side conforms to the covariant transformation given by equation (\ref{newvariation2}). Similar things happen with equation (\ref{in11}). We can expand the first term as
\begin{eqnarray}
 \delta\left[b_{k_1}^{\ \mu}\left(\partial_{\mu} \right.\right.&+&\left.\left. \frac{1}{2}\omega^{ij}_{\mu}\Sigma_{ij}\right)\nabla_{k_2}..\nabla_{k_{n+1}}\phi\right]\nonumber\\
&=& \left(\delta b_{k_1}^{\ \mu}\right)\partial_{\mu}\left(\nabla_{k_2}..\nabla_{k_{n +1}}\phi\right) \nonumber\\&+& b_{k_1}^{\ \mu}\delta \left(\partial_{\mu}\nabla_{k_2}..\nabla_{k_{n +1}}\phi\right)\nonumber\\
&+&\frac{1}{2}\left(\delta b_{k_1}^{\ \mu}\right)\omega^{ij}_{\mu}\Sigma_{ij}\nabla_{k_2}..\nabla_{k_{n+1}}\phi\nonumber\\
&+&\frac{1}{2}b_{k_1}^{\ \mu}\left(\delta\omega^{ij}_{\mu} \right)\Sigma_{ij}\nabla_{k_2}..\nabla_{k_{n+1}}\phi\nonumber\\
&+&\frac{1}{2}b_{k_1}^{\ \mu}\omega^{ij}_{\mu} \Sigma_{ij}\delta\left(\nabla_{k_2}..\nabla_{k_{n+1}}\phi\right)\label{in111}
\end{eqnarray}
Note that the operators $\delta$ and $\partial$ commute. The r.h.s of equation (\ref{in111}) may be computed using equations (\ref{fieldtrans}) and (\ref{newvariation32}). The first term of the r.h.s of (\ref{in11}) thus comes out to be
\begin{eqnarray}
&=&\left(\frac{1}{2}\theta^{ab}\Sigma_{ab} - \xi^\lambda\partial_\lambda\right)\left[b_{k_1}^{\ \mu}\left(\partial_{\mu} +  \frac{1}{2}\omega^{ij}_{\mu}\Sigma_{ij}\right)\nabla_{k_2}..\nabla_{k_{n+1}}\phi\right]\nonumber\\
&+& \theta_{k_1}^{\ m}\left[b_{m}^{\ \mu}\left(\partial_{\mu} +  \frac{1}{2}\omega^{ij}_{\mu}\Sigma_{ij}\right)\nabla_{k_2}..\nabla_{k_{n+1}}\phi\right]\nonumber\\
&+& \theta_{k_2}^{\ m}\left[b_{k_1}^{\ \mu}\left(\partial_{\mu} +  \frac{1}{2}\omega^{ij}_{\mu}\Sigma_{ij}\right)\nabla_{m}..\nabla_{k_{n+1}}\phi\right]\nonumber\\
&+& ....\theta_{k_{n+1}}^{\ m}\left[b_{k_1}^{\ \mu}\left(\partial_{\mu} +  \frac{1}{2}\omega^{ij}_{\mu}\Sigma_{ij}\right)\nabla_{k_2}..\nabla_{m}\phi\right]\nonumber\\
&+& b_{k_1}^{\mu}\partial_{\mu}\theta_{k_2}^{\ m}\nabla_m ...\nabla_{k_{n+1}}\phi + ..\nonumber\\
&+& b_{k_1}^{\mu}\partial_{\mu}\theta_{k_{n+1}}^{\ m}\nabla_{k_2} ...\nabla_{m}\phi\label{1st}
\end{eqnarray}
 Similarly other terms on the r.h.s of (\ref{in11}) may be calculated. We find the second term as 
\begin{eqnarray}
&&\left(\frac{1}{2}\theta^{ab}\Sigma_{ab} - \xi^\lambda\partial_\lambda\right)\left(b_{k_1}^{\ \mu}\omega^s_{\ k_2\mu}\nabla_s...\nabla_{k_{n+1}}\phi\right)\nonumber\\&+& \theta_{k_1}^{\ m}b_{m}^{\ \mu}\omega^s_{\ k_2\mu}\nabla_s...\nabla_{k_{n+1}}\phi
+....+\theta_{k_{n+1}}^{\ m}b_{k_1}^{\ \mu}\omega^s_{\ k_2\mu}\nabla_s...\nabla_{m}\phi\nonumber\\
&+&b_{k_1}^{\ \mu}\partial_{\mu}\theta_{k_2}^{\ m}\nabla_m...\nabla_{k_{n+1}}\label{2nd}
\end{eqnarray}
All terms of (\ref{2nd}) except the last one fit in the covariant structure of (\ref{1st}) while the last term cancels one noncovariant piece of(\ref{1st}). Simlar results follow from the other terms of (\ref{in11}). After simplification of (\ref{in11}) in this way we get
\begin{eqnarray}
\delta\nabla_{k_1}..\nabla_{k_{n+1}}\phi &=& {\cal{P}}\nabla_{k_1}..\nabla_{k_{n+1}}\phi \nonumber\\ &+&\theta_{k_1}^{\ m}\nabla_{m}..\nabla_{k_{n+1}}\phi\nonumber\\&+&...\theta_{k_{n+1}}^{\ m}\nabla_{k_1}....\nabla_{m}\phi 
\label{newvariation321}
\end{eqnarray}
This shows that our assertion (\ref{newvariation32}), if true for $ n$, holds good for $ n+1$ also. So the PGT construction demonstrated for $n=2$ is valid for $n=3$ and so on.

\section{Conclusion}

          We have discussed the extension of Poincare gauge theory framework to matter theories containing higher order derivatives in the lagrangean. The Poincare gauge theory (PGT) \cite{Utiyama:1956sy,Kibble:1961ba} is constructed by localising the Poincare invariance of a matter theory applying the well known gauge principle.  The gauge fields introduced in the process are observed to have one to one correspondence with the geometric structures such as the tetrad and spin -- connection in Riemann -- Cartan spacetime \cite{Blagojevic:2002du,BGMR}. This correspondence enables one to cast gravity in the form of PGT.However, in the usual analysis matter theories limited to lagrangeans containing the first derivative only have been considered . The question of  Poincare gauge theoretic construction from higher derivative matter theory is not much discussed in the literature. Looking from the point of view of the gauge theory, such construction appears to be complicated because it has been found that in general in the extension of the gauge principle to higher derivative theories, new tensor gauge fields are required to be introduced in the process \cite{hama}. On the other hand, the gravitational coupling to the particle theories is usually a straightforward process. The principle of general covariance allows one to couple a Poincare invariant theory to gravity by substituting the Minkowski metric by the metric of curved space -- time and replacing ordinary derivatives by covariant derivative. We have shown by localising Poincare invariance of a higher derivative matter theory that the resulting Poincare gauge theory follows this geometric procedure. Thus our results butresses the geometric interpretation of the PGT observed in the literature from a different perspective. The ordering ambiguity, following from the noncommutativity of the covariant derivative, is manifest in our construction.


\section*{Acknowledgement}

The author would like to thank Rabin Banerjee, Saurav Samanta and Debraj Roy for useful discussion. He also acknowledges the facilities extended to him during his visit to the I.U.C.A.A, Pune and later to S. N. Bose National Centre for Basic Sciences, Kolkata as visiting associate.


\end{document}